\documentclass{article}
\usepackage{amsfonts}
\usepackage{amsmath}
\usepackage[utf8]{inputenc}
\usepackage[super,sort&compress,comma]{natbib} 
\usepackage{graphicx} 
\usepackage{float}
\usepackage[english]{babel}
\addto{\captionsenglish}{%
  
}
\usepackage[T1]{fontenc}
\usepackage{hyperref}

\title{Uniaxial Strain Induced Topological Phase Transition in Bismuth-Tellurohalide--Graphene Heterostructures}

\author{Zoltán Tajkov,\textit{$^{a}$} Dávid Visontai,\textit{$^{b}$} László Oroszlány \textit{$^{c,d}$}\\ and János Koltai\textit{$^{a}$}}

\begin{document}

\maketitle

\noindent\normalsize{We explore the electronic structure and topological phase diagram of heterostructures
formed of graphene and ternary bismuth tellurohalide layers.
We show that mechanical strain inherently present in fabricated samples could
induce a topological phase transition in single-sided heterostructures,
turning the sample into a novel experimental realisation of a time reversal
invariant topological insulator. We construct an effective tight binding
description for low energy excitations and fit the model's parameters to
\emph{ab initio} band structures. We propose a simple approach for predicting
phase boundaries as a function of  mechanical distortions and hence gain a
deeper understanding on how the topological phase in the considered system may
be engineered.} \\






\footnotetext{\textit{$^{a}$ELTE Eötvös Loránd University Department of Biological Physics, Pázmány P. s. 1/A, H-1117, Budapest, Hungary. E-mail: koltai@.elte.hu}}
\footnotetext{\textit{$^{b}$ELTE Eötvös Loránd University Department of Materials Physics, Pázmány P. s. 1/A, H-1117, Budapest, Hungary }}
\footnotetext{\textit{$^{c}$ELTE Eötvös Loránd University Department of Biological Physics, Pázmány P. s. 1/A, H-1117, Budapest, Hungary }}
\footnotetext{\textit{$^{d}$Budapest University of Technology and Economics MTA-BME Lendület Topology and Correlation Research Group, Budafoki út 8., H-1111 Budapest, Hungary }}




\section{Introduction}

In quantum confined nanostructures the electron-spin dephasing time can reach
the order of microseconds, \cite{awschalom_electron_1999,petta_coherent_2005,
hanson_spins_2007} providing an exceptional venue for information processing
and information transmission, such as spin-based devices for conventional
computers or quantum computing.\cite{burkard_spintronics_2000,
leuenberger_spintronics_2001,clemente-juan_magnetic_2012,eremeev_new_2015}
A key difference between spintronic devices and conventional electronics is the
controllable manipulation of the spin degree of freedom of charge carriers.
Manipulating spins without an external magnetic field is necessary for several
technological applications.\cite{zutic_spintronics_2004} Strong spin-orbit
coupling (SOC) in layered two-dimensional (2D) structures potentially leads to
band inversion driving the system through a topological phase
transition.\cite{niu_two-dimensional_2016,ando_topological_2015,
fu_topological_2011}. The created novel quantum phase hosts topologically
protected edge states whose spin is locked to their propagation
direction.\cite{hasan_colloquium_2010} These spin-momentum locked robust edge
states allow for manipulation of the electron spin, hence they have
potential as architectural components in a future quantum information processing
device.\cite{freedman_topological_2003,kitaev_fault-tolerant_2003,
nayak_non-abelian_2008,campbell_roads_2017}

Since its first isolation, graphene emerged as an ideal template material for
revolutionary applications.\cite{novoselov_electric_2004} Although graphene
proved to be an electronic conductor with outstanding mechanical
properties, the SOC is inherently weak in it due to the small atomic weight of
carbon atoms.\cite{tombros_electronic_2007,castro_neto_impurity-induced_2009}
Several theoretical proposals have been made to overcome this limitation for
example with introducing curvature in the graphene sheet or by means of
adatoms.\cite{huertas-hernando_spin-orbit_2006,balakrishnan_colossal_2013} From
an engineering point of view, hybrid 2D heterostructures, due to already existing
and well understood fabrication procedures, seem a more practical approach for
introducing a host of exotic features in graphene samples.\cite{geim_van_2013}
This method has the potential to introduce a considerably large SOC in the
graphene layer of the graphene based devices utilise layered materials with a strong spin-orbit
interaction.

\begin{figure}[h]
\centering
    \includegraphics[width=0.9\textwidth]{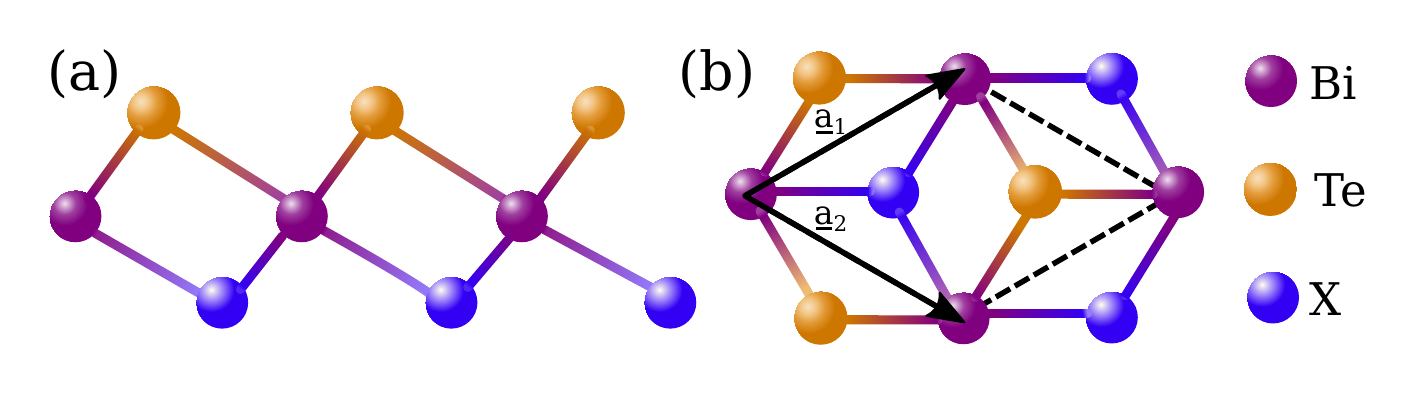}
\caption{%
Side (a) and top (b) view of the structure of a monolayer of the BiTeX
$\mathrm{(X = I,\, Br,\, Cl)}$ crystal. The black dashed line indicates
the unit cell and $\underline{a}_i$ denote the unit cell vectors.}
\label{fig:bitex}
\end{figure}

Ternary bismuth tellurohalides are a new class of polar crystals with a layered
structure represented by BiTeX $\mathrm{(X = I,\,Br,\,Cl)}$.\cite{bahramy_origin_2011}
The key constituent of these compounds is Bi, which is a heavy element and has a
strong atomic SOC. Its triangular lattice layer is stacked between a Te and a
halogen atom layer, see {Figure \ref{fig:bitex}}. The
already large intrinsic SOC in Bi and the structural asymmetry combined with a
large in-plane gradient of the crystal field in this lattice results in giant
Rashba spin-splitting semiconductors.\cite{ast_giant_2007,wu_enhanced_2015,
sakano_three-dimensional_2012,butler_mapping_2014,qi_topological_2017}
Among these BiTeI stands out with the strongest SOC.\cite{ishizaka_giant_2011}
It was also shown that centrosymmetric thin films
composed from topologically trivial
BiTeI trilayers are quantum spin Hall insulators and properly stacked  compounds of BiTeX results in topological insulating phase.
\cite{nechaev_quantum_2017,eremeev_two_2017}
Recently the first experimental isolation and characterisation of a single layer
of BiTeI was reported by F{\"u}l{\"o}p \emph{et al.} using a novel exfoliation
technique.\cite{fulop_exfoliation_2018} Albeit the
band gap increased, the basic characteristics of bulk BiTeI is preserved, hence
this material can be used as a strong SOC inducing component in graphene based
heterostructures, as it was theoretically studied in previous
works.\cite{kou_robust_2014,tajkov_transport_2017}

In this manuscript, we give an account of a detailed theoretical investigation
of heterostructures consisting of graphene and BiTeX layered van der Waals
systems. Based on first principle calculations we distilled an effective model
in order to understand the emerging topological phase in the studied
structures and explored the phase diagram of the systems as a function of
mechanical distortions. We identify single-sided BiTeX-graphene heterostructures
as promising candidates for engineering mechanically controllable topological
phases. {Although mechanical control of topological properties has already been demonstrated, our proposal is the first to point out that uniaxial in-plane strain  can also be used for engineering topological phases. \cite{ren_topological_2016}}

\section{First principles results}

\subsection{Electronic structure of BiTeX-graphene hybrid systems}

\begin{figure}
\centering
\includegraphics[width=0.9\textwidth]{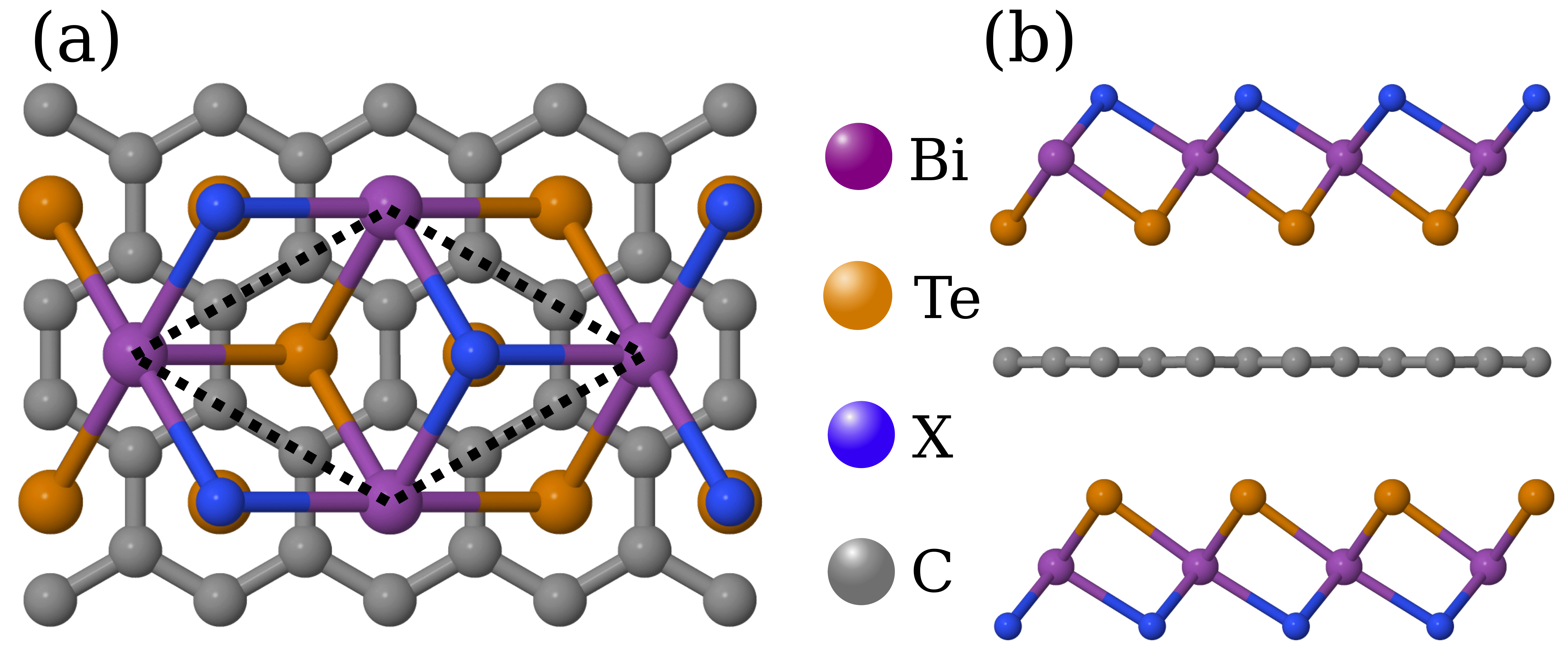}
\caption{\label{fig:szenyo_1}Top (a) and side (b) view of the structure of
graphene/BiTeX (X =I,Br,Cl) sandwich. The black dashed line
indicates the unit cell.}
\end{figure}

The crystal structure of BiTeX is characterised by similar experimental in-plane
lattice constants $a_{\text{BiTeX}}=0.434\,$nm, $\,0.424\,$nm, $0.427\,$nm$\ $ for
$\mathrm{(X = I,\, Br,\, Cl)}$ respectively.\cite{shevelkov_crystal_1995}
As these values are approximately $\sqrt{3}$ times larger than the lattice
constant of graphene ($a_C=0.246\,$nm), it is possible to find a commensurate
supercell by the 30$^\circ$ rotation of the graphene layer.
This supercell is depicted in {Figure \ref{fig:szenyo_1}(a)} consisting
of six carbon atoms, one bismuth, one tellurium and one halogen atom.
Note that this choice leads to a strain not larger than $1.8\%$ in the
BiTeX lattice. {This mismatch may alters the band structure of the BiTeX layer but it does not influence our main conclusions.} Furthermore, we only consider the so called hollow configuration,
that is when the adjacent atom of the BiTeX layer (\textit{i.e.}, Te or X)
is positioned above the centre of a hexagon of carbon atoms in the graphene
layer, as it was shown previously that this horizontal
configuration is the most stable.\cite{kou_robust_2014}

\begin{table}
\center
\caption{\label{table1}
Table of the considered structures. First column denotes the geometry,
where C indicates the graphene layer. The Te layers of BiTeX face graphene,
unless indicated otherwise. Second column are is layer distance
between graphene and BiTeX in nm. The third column shows the band gap calculated by SIESTA in meV, where
"metallic" keyword signifies the absence of a direct band gap. The fourth column
is the value of the $\mathbb{Z}_2$ invariant for insulating systems  {in the
following maner: 0 denotes the trivial and 1 the topological state.}}
\begin{tabular}{c c c c}
\hline
Structure & \shortstack{$d \,$ \\
                        $[$\rm nm$]$}
          & \shortstack{$E_{gap} \,$ \\
                       $[\rm meV]$}
          & $\mathbb{Z}_2$ \\
\hline
BiTeI - C                             & 0.344 & $\approx 1$      & {0}       \\
$2\times$BiTeI - C                    & 0.342 & metallic & N.A.          \\
BiTeI - C [I faced]                   & 0.335 & metallic & N.A.          \\
BiTeI - C - BiTeI                     & 0.345 & 44       & {1}   \\
BiTeI - C - $2\times$BiTeI            & 0.346 & metallic & N.A.          \\
$2\times$BiTeI - C - $2\times$BiTeI   & 0.344 & metallic & N.A.          \\
BiTeCl - C                            & 0.346 & $\approx 1$      & {0}       \\
$2\times$BiTeCl - C                   & 0.344 & metallic & N.A.          \\
BiTeCl - C [Cl faced]                 & 0.300 & metallic & N.A.          \\
BiTeCl - C - BiTeCl                   & 0.346 & 41       & {1}   \\
BiTeCl - C - $2\times$BiTeCl          & 0.344 & metallic & N.A.          \\
$2\times$BiTeCl - C - $2\times$BiTeCl & 0.353 & metallic & N.A.          \\
BiTeBr - C                            & 0.344 & $\approx 1$      & {0}       \\
$2\times$BiTeBr - C                   & 0.339 & metallic & N.A.          \\
BiTeBr - C [Br faced]                 & 0.309 & metallic & N.A.          \\
BiTeBr - C - BiTeBr                   & 0.346 & 42       & {1}   \\
BiTeBr - C - $2\times$BiTeBr          & 0.345 & metallic & N.A.          \\
$2\times$BiTeBr - C - $2\times$BiTeBr & 0.344 & metallic & N.A.          \\
BiTeI - C - BiTeCl                    & 0.344 & 42       & {1}   \\
BiTeI - C - BiTeBr                    & 0.347 & 44       & {1}   \\
BiTeBr - C - BiTeCl                   & 0.344 & 43       & {1}   \\
\hline
\end{tabular}
\end{table}

We compared various combinations of the two constituents  of the investigated
heterostructure, BiTeX layers and graphene. First we studied single-sided
setups, comprising of a single layer of graphene and one or two layers of
BiTeX, taking both Te and X faced alignments into account. We also dealt with
two sided "sandwich" structures, where the graphene sheet is surrounded by one
or two layers of BiTeX from both sides. In the investigated sandwich structures
Te layers were always facing graphene. The initial geometry of all considered
configurations were fully relaxed, allowing also for change of the in-plane
lattice parameter. We found however that in all cases the optimal geometry
yielded the in-plane lattice constant of graphene, as expected due to the larger
stiffness of graphene.\cite{graphene_young_modulus_Lee_2008} The obtained
optimal layer distances, bulk band gap of electronic states (if any) and
topological $\mathbb{Z}_2$ invariant (for insulators) are compiled in
{Table \ref{table1}}.

The layer distance between BiTeX and graphene was found to be around $3.4\,$\AA
$\,$ for all considered arrangements, thus one can safely conclude that the
interlayer interaction in these systems is of van der Waals type. For the
single-sided Br or Cl faced BiTeX graphene configurations the interlayer
distance was found significantly reduced compared to other cases. This tendency
correlates well to the widely accepted, experimentally extracted van der Waals
radii, which are also  definitely lower for Br and Cl atoms compared to Te and
I atoms.\cite{BondivdW1964,RowlandvdW1996}

\begin{figure*}
\centering
\includegraphics[width=0.9\textwidth]{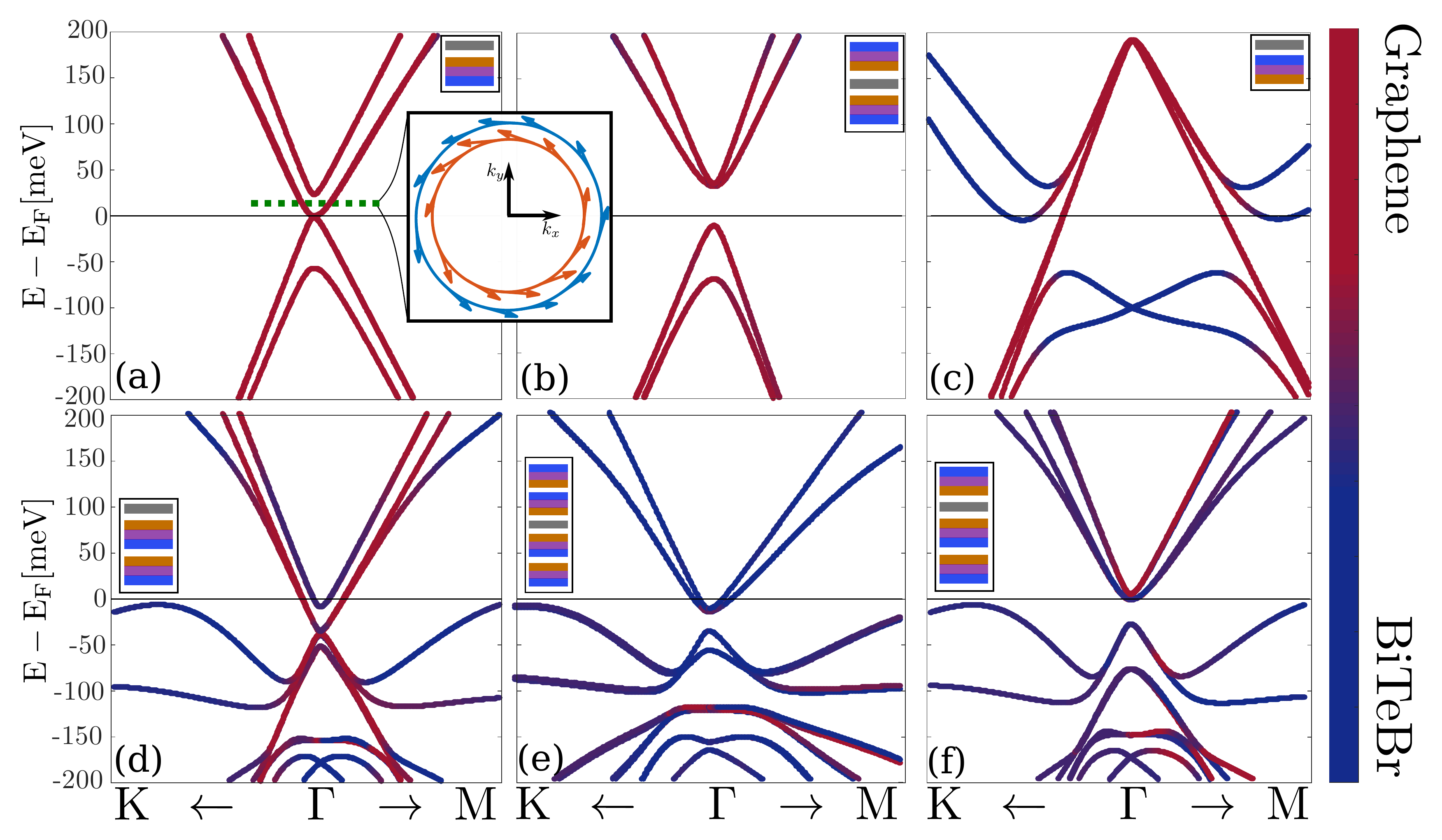}
\caption{
\label{fig:dft_band_structures}
Electronic band structure of the considered geometries consisting of BiTeBr and
graphene layers near the $\Gamma$ point.
Bands are coloured based on the localisation of the states they represent,
hues of red indicate states localised to the graphene layer while blue denotes
BiTeBr orbitals. Insets in all subfigures show the corresponding geometrical
arrangement of atomic layers with the same colour scheme as was introduced in
Figure \ref{fig:szenyo_1}. {The inset between subfigure (a) and (b) shows a schematic representation of the spin direction in a constant energy contour at energy 20 meV (denoted by green dashed line
on the panel). The arrows indicate the direction of the spin expected values in the $k_x$-$k_y$ plane.}}
\end{figure*}

In Figure \ref{fig:dft_band_structures} we present the calculated
electronic band structures for all considered BiTeBr-graphene structures around
the Fermi energy. colouring of the bands corresponds to the orbital weights of
the constituent layers, carbon orbitals are shaded red while states localised to
BiTeBr layers are depicted by blueish colours, a purple hue signifies a strongly
hybridised state.
Depending on the number of BiTeBr layers and their relative alignment
the studied systems show several distinct features.

Figure \ref{fig:dft_band_structures} (a) and (b) depict single-sided and
sandwich structures of Te faced monolayer BiTeBr-graphene arrangements for which
cases the low energy spectrum is dominated by quasiparticle contributions
localised to the carbon atoms, reminiscent of the Dirac cones of graphene.
{In Figure 3(a) the inset shows a schematic representation of the spin texture of the two first conduction bands at a constant energy contour at 20 meV. We note, that in the inset, for clarity, the small but finite spin-splitting of the  lowermost conduction band is artificially enlarged in $k$-space. The arrows representing the direction of the spin expected value in the $x$-$y$ plane. The direction of the arrows are the same in the inner and the outer circle indicating a spin helicity in the graphene Dirac bands. The spin polarization also demonstrates the presence of the out-of-plane spin component. This finding is in good agreement with the observation of Eremeev \emph{et al.}, where the authors examined a similar setup of BiTeCl and graphene.\cite{eremeev_spin-helical_2014}}

 In
these two cases the Dirac point is auspiciously tuned to the gap of the BiTeBr
monolayer bands. Since these structures are primarily characterised by graphene
like bands, they might be understood in terms of a simplified model where the
$p_z$ orbitals of carbon atoms are subject to an induced spin-orbit coupling
(see section \ref{sec:model}).
The main difference between the single-sided and sandwich structures is that,
while in the sandwich structure a sizable direct gap of $41\,\rm{meV}$ is
present at the $\Gamma$ point, in the single-sided system the calculation yields
a considerably smaller gap of about $1\,\rm{meV}$. We also extracted the
$\mathbb{Z}_2$ topological invariant by calculating the flow of Wannier centres
in the half of the Brillouin zone \cite{ryu_topological_2010,asboth_short_2016}
based on the \textit{ab initio} Hamiltonian (for details see section
\ref{methods}). The topological classification of the sandwich structure
confirms the findings of Kou and coworkers namely that the gap in sandwich
structures is topological.\cite{kou_robust_2014} On the other hand based on our
calculations the small gap in the single-sided system is trivial.

In the other investigated scenarios depicted in Figure \ref{fig:dft_band_structures}
(c)-(f), in contrast to the Te faced monolayer BiTeBr-graphene arrangements,
the image of the Dirac point is either considerably shifted away
from the Fermi level or it is masked by states originating from BiTeBr, thus
a strong mixing of BiTeBr and carbon bands occurs, resulting in metallic
band structures. Similar trends can be identified in other arrangements,
independently from the type of halogen atom considered (cf. Table \ref{table1}).

In the  {remainder} of the manuscript we shall focus on the setups where the low
energy electronic structure is dominated by the $p_z$ orbitals of graphene.
Therefore, we shall only consider the Te faced sandwich and the monolayer
BiTeX-graphene arrangements.

\subsection{Effect of mechanical distortions}
\begin{figure}
\centering
    \includegraphics[width=0.9\textwidth]{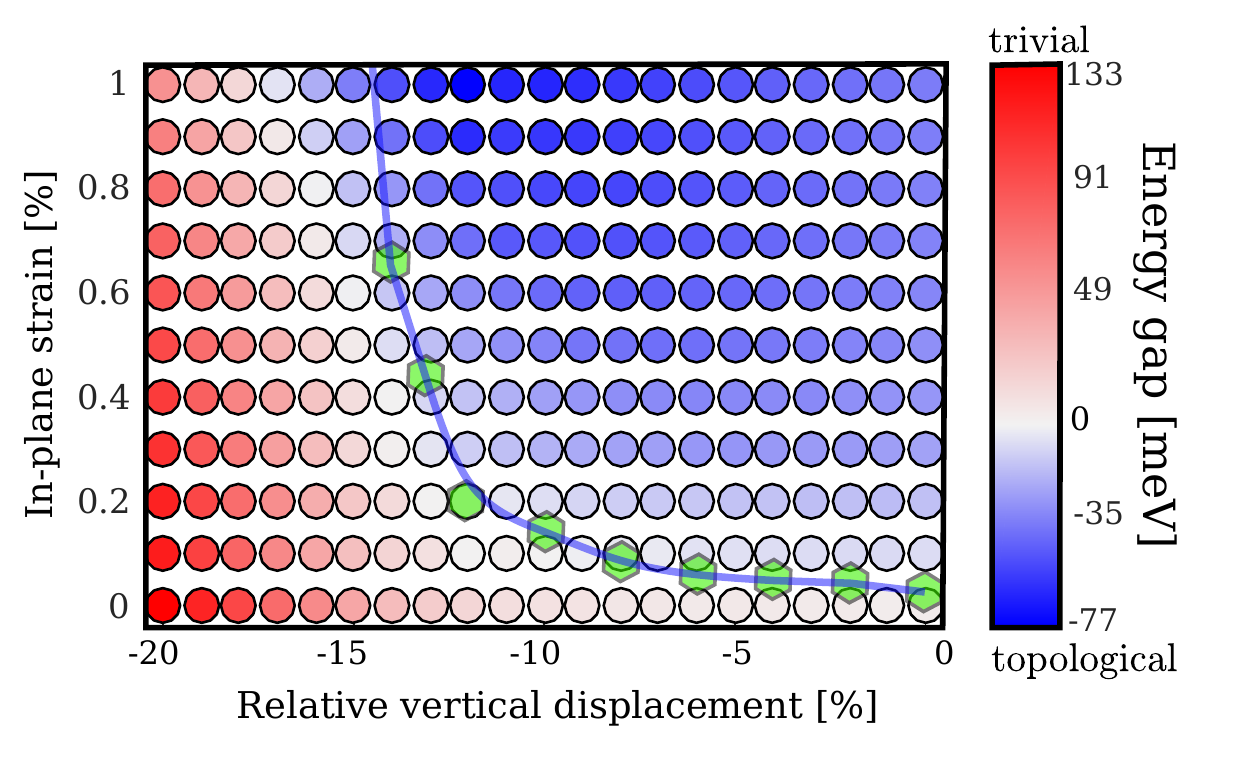}
\caption{%
The circles show the \emph{ab initio} phase diagram of the single-sided
BiTeBr-graphene heterostructure in the presence of mechanical distortion.
Topological band gaps are denoted by negative values.
Green hexagons (connected with a blue tentative curve) trace out the tight binding
phase boundary (see text for more details).
\label{fig:dft_phase}}
\end{figure}

Previously, we presented results obtained for relaxed geometries. Manufactured
devices are commonly subject to mechanical distortions which in turn can have a
non-negligible impact on the electronic properties of the system. Recently
considerable experimental effort has been made to make use of strain fields to
control electronic and optical properties of novel heterostructures.
\cite{CastellanosGomez_Strain_Engineering_MoS2_2013,Jiang_Pseudomagnetic,
Sanchez_strain_tuning_2017,Goldsche_Strain_Fields_graphene_2018}
Inspired by these advances, we investigate below how the electronic states
and their topological character is influenced in Te faced single-sided
BiTeBr-graphene heterostructure as the sample is subjected to in-plane uniaxial
strain combined with stress perpendicular to the device.

We model in-plane uniaxial strain in our \emph{ab initio} calculations by
stretching the in-plane unit cell vectors along a carbon--carbon bond, and
allowing the atomic positions relax in the constrained unit cell.
On the other hand out-of-plane strain is simulated by reducing the
distance between graphene and the BiTeX layer, without relaxation of the atomic
positions.

We calculated the evolution of the size of the band gap and the topological
index as both in-plane and out-of-plain strain was varied. The corresponding
phase diagram is shown on {Figure \ref{fig:dft_phase}}. The sign of the
band gap indicates the topological invariant, it is negative if the system is
topological  {and} positive otherwise.

Based on the presented results we conclude that both type of mechanical
distortions have a striking effect on the band gap, however they favour
different topological phases. Out-of-plane strain widens the initially present
trivial band gap, while in-plane strain drives the system {into}
the topological phase. We only present results for BiTeBr further calculations show
that BiTeCl behaves qualitatively in the same fashion. BiTeI turns metallic
instead of a trivial insulator due to some non graphene bands reaching the Fermi
level as pressure is increased.

We note that the largest out-of plane strain we applied would correspond
to a nominal pressure of $20\,$GPa, which we estimated as the energy derivation
per unit area over the reduced distance. The mechanical stresses considered in
our calculations can thus be routinely achieved in nowadays experimental
setups.\cite{song_largely_2018,strain_1,strain_2,kleppe_new_2014,ni_uniaxial_2008,braganza_hydrostatic_2002,
vos_high-pressure_1992,kullmann_effect_1984}

\section{\label{sec:model}Low energy description}

\subsection{A model Hamiltonian}

\begin{figure}
\centering
    \includegraphics[width=0.9\textwidth]{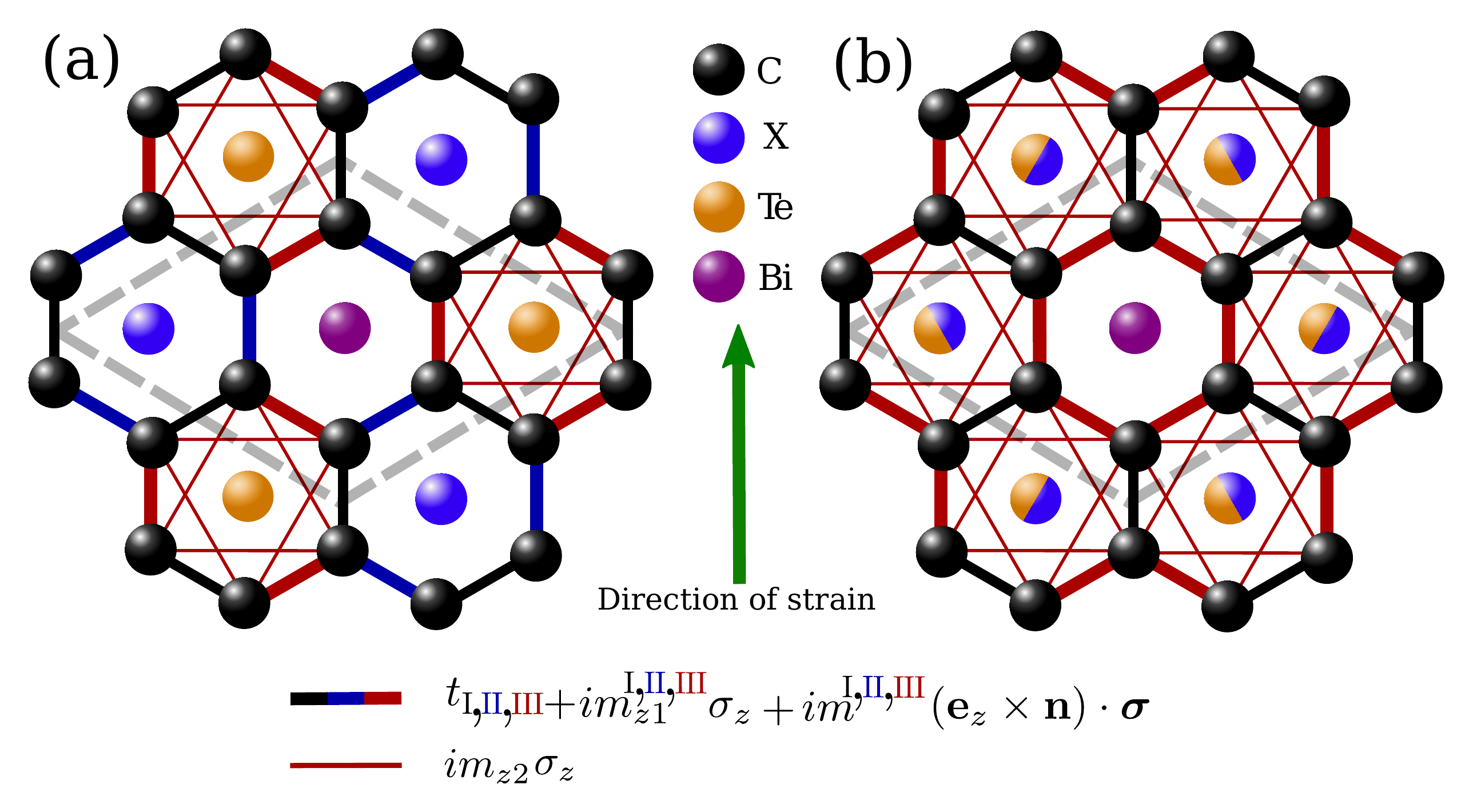}
\caption{\label{fig:model_geom}
Illustration of the tight binding model of the graphene/BiTeX heterostructures for
single-sided (a) and sandwich (b) arrangements. {The alternating colouring
of X and Te atoms in inset (b) indicates the inversion symmetry of the system}}
\end{figure}


In the previous section, based on first principles calculations, we identified
two arrangements from all the considered structures where electronic states
close to the Fermi energy are dominated by the $p_z$ orbitals of carbon atoms.
These two setups, the Te faced single-sided BiTeX-graphene heterostructure
and the sandwich system, are depicted in the insets of Figure
\ref{fig:dft_band_structures} (a)  and (b).
In this section we propose an effective model, based on the tight binding (TB)
description of graphene, where due to the presence of BiTeX layers an
appropriately patterned SOC emerges.
With the help of the introduced model we gain a deeper insight into the
mechanisms responsible for the emergence of the topological phase witnessed
before. Our effective description of the heterostructure is cast in a form that
is digestible for theoretical approaches calculating electron, thermal and
spin transport properties of samples on experimentally relevant
scales.\cite{equus,Ferrer_Gollum_2014}

Respecting the symmetries of the system studied in the first principles
calculations, we introduce a model with only a minimal number of parameters.
Beyond the usual hopping integrals $t$, we assume spin-orbit interaction
compatible with time reversal symmetry on nearest neighbour carbon-carbon bonds
as $i\mathbf{m} \boldsymbol{\sigma}$, where $\mathbf{m}$ is a real vector,
$\boldsymbol{\sigma}$ is the vector of Pauli matrices. We also include second
nearest neighbour out-of-plane spin-orbit interaction as it was previously
proposed by Kane and Mele to germinate a topological phase transition in
graphene like systems.\cite{Kane_Mele}
{The studied systems are invariant under the symmetries of the $C_{3v}$ point group.}
The in-plane component of the $\mathbf{m}$ vector on each bond is restricted
by this symmetry and points perpendicular to the bonds \cite{winkler_extended_2003}.
After taking into account all the symmetries of the system our simplified model
is depicted on {Figure \ref{fig:model_geom}a}. {Thus we} introduce three
different nearest neighbour hopping integrals $t_{\mathrm{I,II,III}}$, three
in-plane $m^{\mathrm{I,II,III}}$ and three out-of-plane
$m^{\mathrm{I,II,III}}_{z1}$ spin-orbit interaction magnitudes and a
second-nearest neighbour out-of-plane spin-orbit interaction strength
$m_{z2}$.

In the absence of the BiTeX layers graphene also possesses inversion symmetry
with an inversion centre in the middle of each hexagon. This symmetry is
altered in the case of the sandwich arrangement in such a way that only hexagon
centres aligned with Bi atoms remain inversion centres. In this case inversion
symmetry forbids SOC on first nearest-neighbour bonds (denoted by thick black
lines in Figure \ref{fig:model_geom}b), while it connects the hopping and SOC
magnitudes along bonds {that was marked by red and blue thick lines in Figure
\ref{fig:model_geom}a), that is the two hoppings must be equal in the case of
inversion symmetry. This yields}
 $t_{\mathrm{II}}=
t_{\mathrm{III}}$, $m^{\mathrm{II}}=m^{\mathrm{III}}=m$ and $m^{\mathrm{II}}_{z1}=
m^{\mathrm{III}}_{z1}=m_{z1}$ while $m^{\mathrm{I}}=m^{\mathrm{I}}_{z1}=0$.
We now make some further assumptions regarding the actual investigated system.
Since the interaction between BiTeX and graphene is van der Waals type and the
Bi atoms are further away from the graphene sheet as the Te atoms we neglect the
second nearest neighbour SOC on the hexagonal plaquettes encircling Bi atoms.

Breaking inversion symmetry by separating one of the BiTeX layer from the sandwich
structure results in a Te faced single-sided geometry. In this case nearest
neighbour SOC is allowed on thick black bonds, and blue and red bonds are no
longer connected. Consistently with our previous approach we neglect second
nearest neighbour SOC on the plaquettes encircling X atoms since they are even
further from the graphene sheet as Bi. Therefore the single-sided setup is
characterised by 10 different parameters, the hopping amplitudes $t_i$,
first nearest neighbour in-plane SOC $m^i$ and out-of plane SOC $m^i_{z1}$ with $i=
\mathrm{I,II,III}$ and the only second nearest neighbour SOC $m_{z2}$ in hexagons
surrounding Te atoms. {In both cases the red thin lines denote the considered second nearest neighbour SOC interactions in Figure \ref{fig:model_geom}.}

We thus constructed the {real space} Hamiltonian as:

\begin{equation}
\label{hamiltonian}
\begin{split}
H =\sum_{\alpha}\sum_{\substack{p=1,3,5 \\ q=2,4,6}}\left[ \left(
t_{\mathrm{II}}\sigma_0+im^{\mathrm{II}}_{z1}\sigma_z +im^{\mathrm{II}} \left(
\mathbf{e}_z \times \mathbf{n}_{\alpha\alpha pp+1 }\right)\cdot \boldsymbol{
\sigma} \right)\hat{c}^{\dagger}_{\alpha p}\hat{c}_{\alpha p+1} \right. \\
+\left. \left( t_{\mathrm{III}}\sigma_0+im^{\mathrm{III}}_{z1}\sigma_z +im^{
\mathrm{III}} \left(\mathbf{e}_z \times \mathbf{n}_{\alpha\alpha qq+1 }\right)
\cdot \boldsymbol{\sigma} \right)  \hat{c}^{\dagger}_{\alpha q}\hat{c}_{\alpha
q+1}\right] \\
+\sum_{\substack{<\alpha \gamma> \\ <pq>}}\left( t_{\mathrm{I}}\sigma_0+
im^{\mathrm{I}}_{z1}\sigma_z +im^{\mathrm{I}} \left(\mathbf{e}_z \times
\mathbf{n}_{\alpha\gamma pq }\right)\cdot \boldsymbol{\sigma} \right)
\hat{c}^{\dagger}_{\alpha p}\hat{c}_{\gamma q} \\
+\sum_{\substack{<\alpha \gamma>\\ \ll pq \gg }}im_{z2}\sigma_z
\hat{c}^{\dagger}_{\alpha p}\hat{c}_{\gamma q}+h.c..
\end{split}
\end{equation}

The sum in $\alpha$ and $\gamma$ goes over all unit cells in the crystal, while
$p,q$ indicate one of the six atoms in a given unit cell.
Annihilation (creation) operators for an electron in unit cell $\alpha$ on site
$p$ are denoted by $\hat{c}^{(\dagger)}_{\alpha p}$, $\mathbf{e}_z$ is a unit
vector pointing in the $z$ direction, $\mathbf{n}_{\alpha\gamma p q}$ is the
vector that points from site $i$ in unit cell $\alpha$ to site $j$ in unit cell
$\gamma$ and $\sigma_0$ is the $2\times 2$ identity operator.

\begin{figure}
  \centering
  \includegraphics[width=0.9\textwidth]{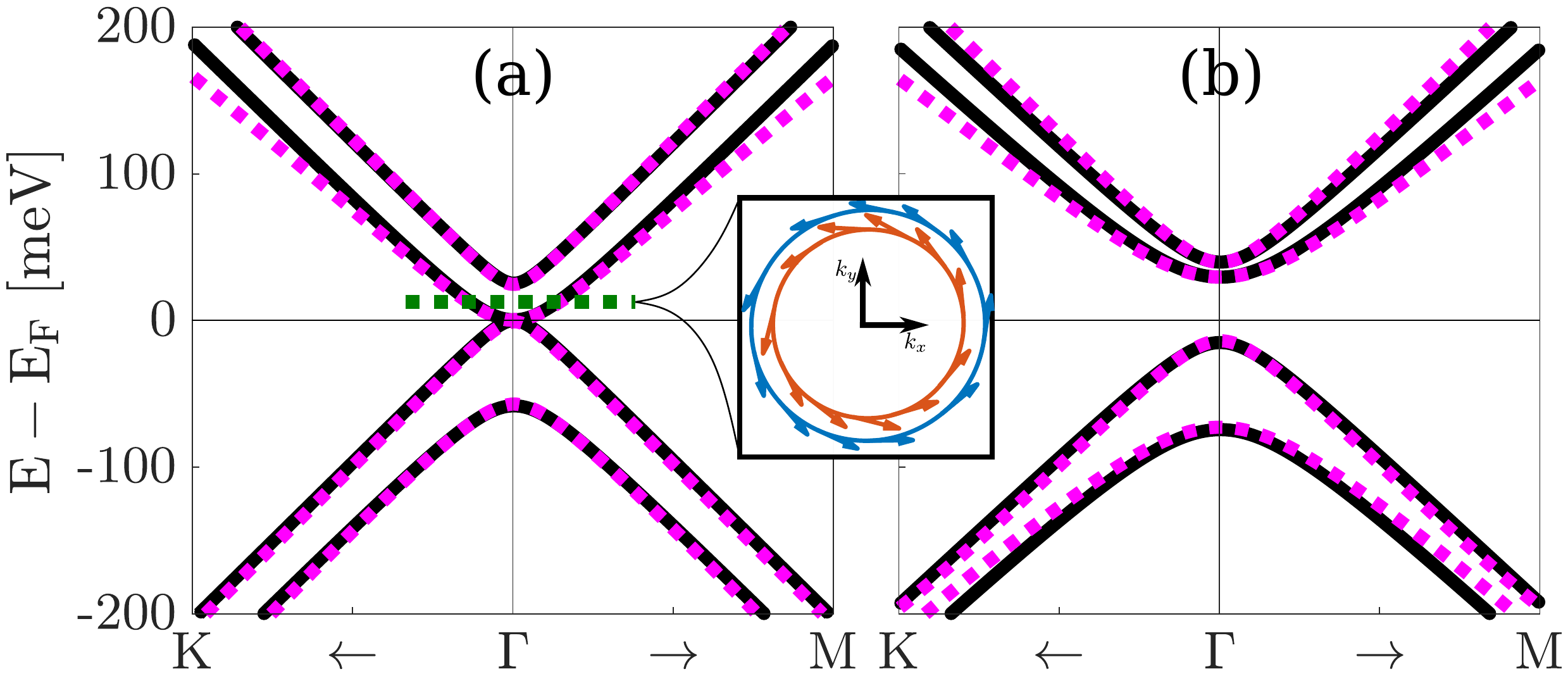}
\caption{\label{fig:fitted_band_structures}The band structure of the single-sided (a)
and the sandwich (b) BiTeBr-graphene heterostructure in the low energy range near
the $\Gamma$-point. Solid black lines indicate the first principles result and dotted
magenta lines mark our fitted tight binding bands. For the fitted parameters see
{Table \ref{table2}}. {The inset between subfigure (a) and (b) shows the 
schematic representation of the spin-texture of the fitted single sided graphene--BiTeBr
heterostructure on a constant energy contour at 20 meV (denoted by green dashed line
on the panel.)}}
\end{figure}

\begin{figure}
\centering
    \includegraphics[width=0.9\textwidth]{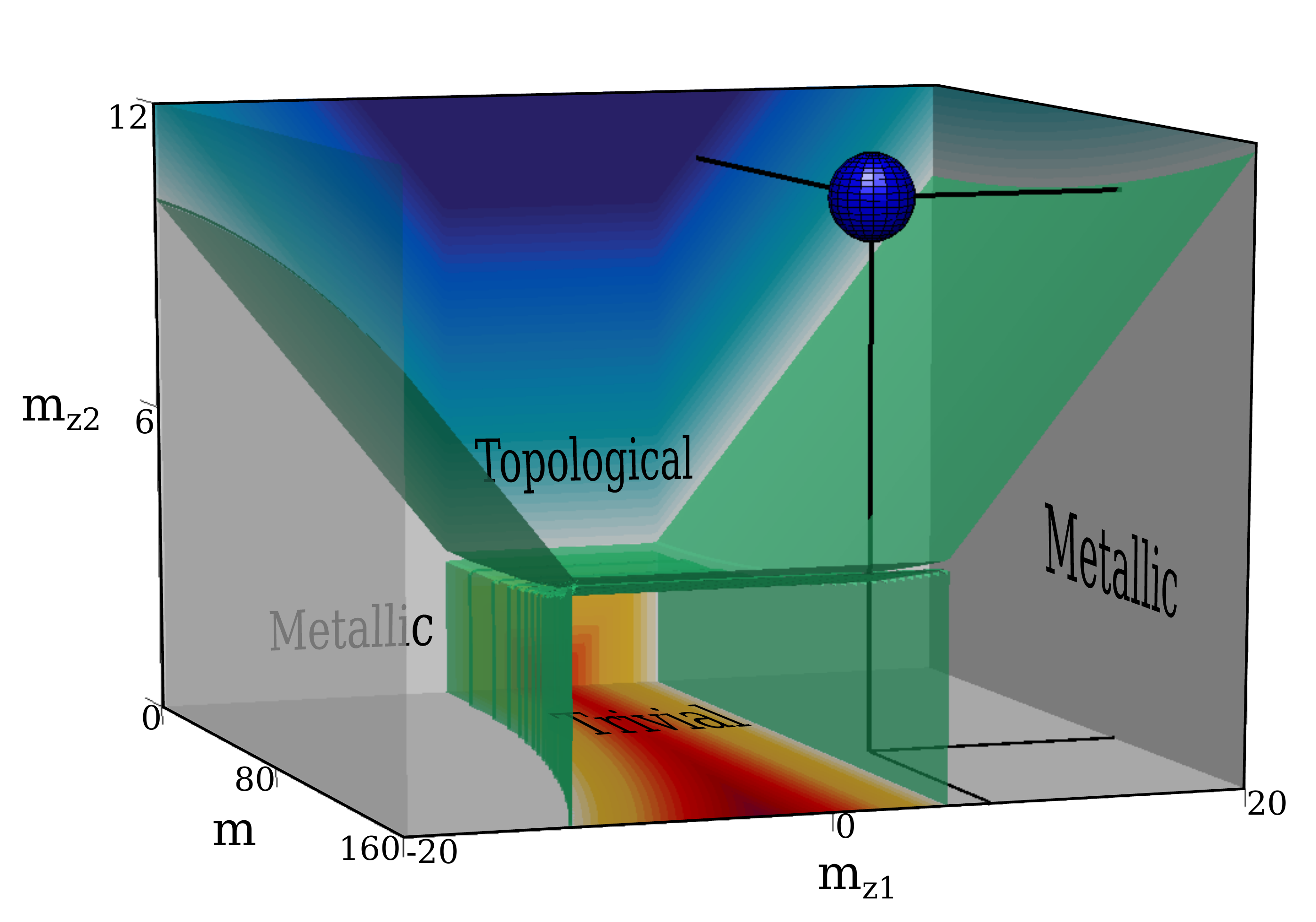}
\caption{\label{fig:phase_of_model}
Topological phase diagram of the inversion symmetric graphene/BiTeBr structure
described by {Equation (\ref{hamiltonian})} in the case of $t_{\mathrm{I}} =
2.227\,\mathrm{eV}$ and $t_{\mathrm{II}} = 2.210\,\mathrm{eV}$.
The $m$, $m_{z1}$ and $m_{z2}$ parameters are shown in units of meV. {The blue sphere pins where the fitted parameters land in the phase diagram.}}
\end{figure}

The fitted and first principles band structure are depicted together for both
the single-sided and sandwich structures in
{Figure \ref{fig:fitted_band_structures}}.
In the following we map the topological character of the introduced two models.
Tracking again the evolution of Wannier centres yields the $\mathbb{Z}_2$
topological invariant for the proposed effective
Hamiltonian.\cite{ryu_topological_2010,asboth_short_2016}
First we investigate the inversion symmetric sandwich setup, and focus on the
parameters responsible for spin-orbit interaction, $m$, $m_{z1}$ and $m_{z2}$
while fixing the two hopping terms to $t_{\mathrm{I}} = 2.227\,\mathrm{eV}$
and $t_{\mathrm{II}} = 2.210\,\mathrm{eV}$.
The phase diagram in {Figure \ref{fig:phase_of_model}} shows two
topologically distinct insulating phases and a metallic phase.
We observe that without SOC but in the presence of a hopping mismatch the model
is a trivial Kekulé patterned band insulator as it was studied in clean graphene
under appropriate biaxial strain.\cite{Spontanious_Kekule_graphene} The first
nearest neighbour out-of plane SOC $m_{z1}$ drives the system towards a gapless metallic
phase, while the second nearest neighbour SOC $m_{z2}$ promotes the topological insulator
phase, large $m$ eventually {drives the system into the} trivial insulator {phase} as it also does in more
simple models.\cite{Kane_Mele}
A blue sphere denotes the parameter configuration that represent the fit to
the \emph{ab initio} result for the sandwich configuration.
\begin{table*}[h!]
\center
\caption{Table of the fitted tight binding parameters (meV) as defined in the Hamiltonian Equation (\ref{hamiltonian})}
\scalebox{0.752}{\begin{tabular}{c c c c c c c c c c c }
\hline
Structure & $t_{\mathrm{I}}$ & $t_{\mathrm{II}}$ & $t_{\mathrm{III}}$ & $m^{\mathrm{I}}$ & $m^{\mathrm{II}}$ & $m^{\mathrm{III}}$ & $m^{\mathrm{I}}_{z1}$ & $m^{\mathrm{II}}_{z1}$ &$m^{\mathrm{III}}_{z1}$ & $m_{z2}$ \\
\hline
BiTeI - C  & {2400} & {2580}  &  {2369} & {-0.232} & {-403.1} & {-396.3} & {0.010} & {28.97} & {-3.797} & {-11.80}  \\
BiTeBr - C & {2379} & {2382}  &  {2376} & {-0.586} &  {147.3} &  {188.9} &  {3.227} &  {4.434} & {7.533} & {-5.722}  \\
BiTeCl - C & {2359} & {2369}  &  {2370} & {0.568} & {-29.95} &  {-38.77} &  {0.061} &  {20.89} & {-5.685} & {-11.59}  \\
BiTeI - C - BiTeI   & 2268 & \multicolumn{2}{c}{2274} & 0 & \multicolumn{2}{c}{146.9} & 0 & \multicolumn{2}{c}{7.304} & 11.15  \\
BiTeBr - C - BiTeBr & 2227 & \multicolumn{2}{c}{2210} & 0 & \multicolumn{2}{c}{138.4} & 0 & \multicolumn{2}{c}{11.90} & 10.64  \\
BiTeCl - C - BiTeCl & 2251 & \multicolumn{2}{c}{2258} & 0 & \multicolumn{2}{c}{129.8} & 0 & \multicolumn{2}{c}{6.040} & 9.900  \\
\hline
\end{tabular}}
\label{table2}
\end{table*}

\subsection{Understanding the role of mechanical distortions}

Turning our attention to the single-sided arrangements our main goal is to
understand the phase diagram emerging from the first principles calculations
presented in Figure \ref{fig:dft_phase} describing a topological phase
transition driven by in-plane uniaxial strain and compression perpendicular to the
sample. In the absence of in-plane distortions but under out-of-plane strain
the system can still be well approximated by the model described in the previous
section. Unavoidably, the application of uniaxial deformation breaks $C_{3v}$
symmetry of the system. We included this effect into our model by multiplying
all first nearest neighbour hopping and SOC terms in the direction of the
uniaxial strain by a common factor $\mathrm{e}^{-\beta(l/l_0-1)}$.\cite{papaconstantopoulos_tight-binding_1997} In this
multiplicative term $l_0$ is the bond length in stress free system, $l$ is the
bond length in the deformed case and $\beta=2.384$ is a dimensionless parameter
determined from fitting to \emph{ab initio} calculations performed on freestanding
graphene. This approach is widely used in the literature for describing
the effects of mechanical distortions on TB parameters.
\cite{papaconstantopoulos_tight-binding_1997,guinea_strain_2012}
We note that following common practice bonds not parallel with the distortion
were not modified. We also remark that this method is applicable for small to
moderate strain values where buckling of the hexagons is negligible.
\cite{rezaei_modified_2018}

Using this simple model for in-plane distortions we are now equipped for
exploring the phase diagram of the effective system.
We first fit our $C_{3v}$ symmetric model to \emph{ab initio} calculations performed
for geometries without uniaxial in-plane strain. Introducing strain in these
models as described above we calculate the size of the quasi particle gap and
the topological invariant. The phase boundary of the TB model
is depicted with green hexagons in Figure \ref{fig:dft_phase}. It shows that
for all cases the initially present trivial gap closes and a topological gap
opens as one tunes the magnitude of uniaxial strain $(l/l_0-1)$ in the multiplicative factor. 
As it can be observed in Figure \ref{fig:dft_phase}. our rudimentary approach
for including the strain in the effective model adequately predicts the boundary
of the topological phase, even for moderately large compressions, thus justifying its application.
One can understand the emerging topological phase transition in the
investigated heterostructures as a competition of a trivial gap of Kekulé type
and a topological Kane-Mele type.\cite{Spontanious_Kekule_graphene,Kane_Mele}
Due to time reversal symmetry uniaxial strain displaces the Dirac cones of
graphene in the two valleys in the opposite direction. Since Kekulé distortion
hybridises the different valleys it is impeded by the strain.
At a given finite strain value the topological component thus overpowers the
trivial and a system turns into a time reversal invariant topological insulator.

\section{Methods}
\label{methods}
The optimised geometry and ground state Hamiltonian and overlap matrix elements
of each structure were self consistently obtained by the SIESTA implementation
of density functional theory (DFT)\cite{noauthor_siesta_2002,artacho_siesta_2008}.
SIESTA employs norm-conserving pseudopotentials to account for the core
electrons and linear combination of atomic orbitals to construct the valence
states. For all cases the considered samples were separated with a minimum of $1.85\,$nm
$\ $ thick vacuum in the perpendicular direction. The generalised gradient
approximation of the exchange and the correlation functional was used with
Perdew-Burke-Ernzerhof parametrisation\cite{perdew_generalized_1996} and
the pseudopotentials optimised by Rivero \textit{et al}.\cite{rivero_systematic_2015}
with a double-$\zeta$
polarised basis set and a real-space grid defined with an equivalent energy
cutoff of $1000\,\rm Ry$. The Brillouin zone integration was sampled by a
$24 \times 24\times 1$ Monkhorst-Pack $k$-grid.\cite{monkhorst_special_1976}
The geometry optimisations were performed until the forces were smaller than
$0.1\, \mathrm{eV}/$nm. The choice of {pseudopotentials} optimised by Rivero
\textit{et al}. ensures that both the obtained geometrical structures and the
electronic band properties are reliable. As a benchmark we validated our method
by comparing the electronic properties of the 
bulk BiTeI with the experimental data. This approach gave us $130\, \rm meV$ band gap and $0.46\, \rm eVnm$ as Rashba parameter for BiTeI bulk. The corresponding experimental results are $130\, \rm meV$ and $0.38\, \rm eVnm$ respectively. \cite{ishizaka_giant_2011} {Relativistic effects, including spin-orbit coupling, were fully taken into account in every performed calculation. \cite{fernandez-seivane_site_2006}}

The fitting procedure was carried out by applying a constrained least squares
minimisation procedure to the difference between the TB and the DFT band energies.
As one  progresses  further  in  energy  away  from  the  Fermi level
the assumption that carbon $p_z$ orbital contributions dominate begins to break down,
with significant BiTeX orbital contributions at energies around $\pm 300\,\rm meV$.
Therefore we fit the model to only 8 bands in an energy window of $\pm 200\,\rm meV$.
The fit is carried out over 720 points in $\mathbf{k}$-space along the high symmetry
lines of the Brillouin zone. {While the procedure is in principle straightforward, in practice one must take care, in particular with the choice of bands to use for the fitting procedure. On diagonalization the model yields 12 bands. We fit the model to the 4 highest energy occupied and the 4 lowest energy valence bands, as the graphene’s Dirac cones are 8 times degenerate in the gamma point in the case of a $\sqrt{3} \times \sqrt{3}$ supercell. Additionally we took into account the $\mathbb{Z}_2$ invariant of the system and the spin structure of the 8 fitted bands, that is calculating the expected values of the spin orientations. The fitness of the procedure is presented in the inset figure of Figure \ref{fig:fitted_band_structures}, which is in good agreement of the DFT results (inset figure of Figure \ref{fig:dft_band_structures})}.

\section{Conclusion}

In summary, we explored the rich topological phase diagram of bismuth
tellurohalide/graphene heterostructures by means of first principles calculations.
Based on our \emph{ab initio} results we distilled a simple tight binding description
for the investigated system capturing all relevant features of the low energy
spectra of quasiparticles. We have demonstrated that the topological phase transition due
to mechanical distortions in one-sided systems leads to a novel realisation of the
time reversal invariant topological insulating phase, thus making these heterostructures
potential candidates for quantum technology applications.

\section*{Conflicts of interest}
There are no conflicts to declare.

\section*{Acknowledgements}

This work was financially supported by the  the Hungarian National Research,
Development and Innovation Office (NKFIH) via the National Quantum Technologies
Program 2017-1.2.1-NKP-2017-00001; grants no. K112918, K115608, FK124723,
K115575 and Flag-ERA iSpinText project NN118996. This work was completed in the ELTE Excellence Program
(1783-3/2018/FEKUTSTRAT) supported by the Hungarian Ministry of Human Capacities.
LO, JK acknowledge the Bolyai and Bolyai+ program of the Hungarian Academy of
Sciences. We acknowledge [NIIF] for awarding us access to resource based in
Hungary at Debrecen.

\bibliography{cikkek} 
\bibliographystyle{unsrt} 

\end{document}